\documentstyle[11pt,psfig,twoside]{article}

\begin{document}

\newcommand{\dd}{\,{\rm d}}
\newcommand{\ie}{{\it i.e.},\,}
\newcommand{\etal}{{\it et al.\ }}
\newcommand{\eg}{{\it e.g.},\,}
\newcommand{\cf}{{\it cf.\ }}
\newcommand{\vs}{{\it vs.\ }}
\newcommand{\zdot}{\makebox[0pt][l]{.}}
\newcommand{\up}[1]{\ifmmode^{\rm #1}\else$^{\rm #1}$\fi}
\newcommand{\dn}[1]{\ifmmode_{\rm #1}\else$_{\rm #1}$\fi}
\newcommand{\upd}{\up{d}}
\newcommand{\uph}{\up{h}}
\newcommand{\upm}{\up{m}}
\newcommand{\ups}{\up{s}}
\newcommand{\arcd}{\ifmmode^{\circ}\else$^{\circ}$\fi}
\newcommand{\arcm}{\ifmmode{'}\else$'$\fi}
\newcommand{\arcs}{\ifmmode{''}\else$''$\fi}
\newcommand{\MS}{{\rm M}\ifmmode_{\odot}\else$_{\odot}$\fi}
\newcommand{\RS}{{\rm R}\ifmmode_{\odot}\else$_{\odot}$\fi}
\newcommand{\LS}{{\rm L}\ifmmode_{\odot}\else$_{\odot}$\fi}

\newcommand{\Abstract}[2]{{\footnotesize\begin{center}ABSTRACT\end{center}
\vspace{1mm}\par#1\par
\noindent
{\bf Key words:~~}{\it #2}}}

\newcommand{\TabCap}[2]{\begin{center}\parbox[t]{#1}{\begin{center}
  \small {\spaceskip 2pt plus 1pt minus 1pt T a b l e}
  \refstepcounter{table}\thetable \\[2mm]
  \footnotesize #2 \end{center}}\end{center}}

\newcommand{\TableSep}[2]{\begin{table}[p]\vspace{#1}
\TabCap{#2}\end{table}}

\newcommand{\TableFont}{\footnotesize}
\newcommand{\TableFontIt}{\ttit}
\newcommand{\SetTableFont}[1]{\renewcommand{\TableFont}{#1}}

\newcommand{\MakeTable}[4]{\begin{table}[htb]\TabCap{#2}{#3}
  \begin{center} \TableFont \begin{tabular}{#1} #4 
  \end{tabular}\end{center}\end{table}}

\newcommand{\MakeTableSep}[4]{\begin{table}[p]\TabCap{#2}{#3}
  \begin{center} \TableFont \begin{tabular}{#1} #4 
  \end{tabular}\end{center}\end{table}}

\newenvironment{references}%
{
\footnotesize \frenchspacing
\renewcommand{\thesection}{}
\renewcommand{\in}{{\rm in }}
\renewcommand{\AA}{Astron.\ Astrophys.}
\newcommand{\AAS}{Astron.~Astrophys.~Suppl.~Ser.}
\newcommand{\ApJ}{Astrophys.\ J.}
\newcommand{\ApJS}{Astrophys.\ J.~Suppl.~Ser.}
\newcommand{\ApJL}{Astrophys.\ J.~Letters}
\newcommand{\AJ}{Astron.\ J.}
\newcommand{\IBVS}{IBVS}
\newcommand{\PASP}{P.A.S.P.}
\newcommand{\Acta}{Acta Astron.}
\newcommand{\MNRAS}{MNRAS}
\renewcommand{\and}{{\rm and }}
\section{{\rm REFERENCES}}
\sloppy \hyphenpenalty10000
\begin{list}{}{\leftmargin1cm\listparindent-1cm
\itemindent\listparindent\parsep0pt\itemsep0pt}}%
{\end{list}\vspace{2mm}}

\def\TYLDA{~}
\newlength{\DW}
\settowidth{\DW}{0}
\newcommand{\dw}{\hspace{\DW}}

\newcommand{\refitem}[5]{\item[]{#1} #2%
\def\REFARG{#3}\ifx\REFARG\TYLDA\else, {\it#3}\fi
\def\REFARG{#4}\ifx\REFARG\TYLDA\else, {\bf#4}\fi
\def\REFARG{#5}\ifx\REFARG\TYLDA\else, {#5}\fi.}

\newcommand{\Section}[1]{\section{#1}}
\newcommand{\Subsection}[1]{\subsection{#1}}
\newcommand{\Acknow}[1]{\par\vspace{5mm}{\bf Acknowledgements.} #1}
\pagestyle{myheadings}

\def\thefootnote{\fnsymbol{footnote}}

\begin{center}
{\Large\bf The Optical Gravitational Lensing Experiment.\\
\vskip3pt
BVI Maps of Dense Stellar Regions.\\
\vskip3pt
I. The Small Magellanic Cloud\footnote{Based on observations obtained with the 
1.3~m Warsaw telescope at the Las Campanas Observatory of the Carnegie 
Institution of Washington.}} 
\vskip1cm

\def\thefootnote{\fnsymbol{footnote}}
{\bf A.~~U~d~a~l~s~k~i$^1$,~~M.~~S~z~y~m~a~{\'n}~s~k~i$^1$,~~
M.~~K~u~b~i~a~k$^1$,\\ 
G.~~P~i~e~t~r~z~y~\'n~s~k~i$^1$,~~ 
P.~~W~o~\'z~n~i~a~k$^2$,~~ and~~K.~~\.Z~e~b~r~u~\'n$^1$}
\vskip5mm
{$^1$Warsaw University Observatory, Al.~Ujazdowskie~4, 00-478~Warszawa, Poland\\
e-mail: (udalski,msz,mk,pietrzyn,zebrun)@sirius.astrouw.edu.pl\\
$^2$ Princeton University Observatory, Princeton, NJ 08544-1001, USA\\
e-mail: wozniak@astro.princeton.edu}
\end{center}
\vskip 10mm

\Abstract{We present three color, {\it BVI} maps of the Small Magellanic 
Cloud. The maps contain precise photometric and astrometric data for about 2.2 
million stars from the central regions of the SMC bar covering ${\approx2.4}$ 
square degrees on the sky. Mean brightness of stars is derived from 
observations collected in the course of the OGLE-II microlensing search from 
about 130, 30 and 15 measurements in the {\it I}, {\it V} and {\it B}-bands, 
respectively. Accuracy of the zero points of photometry is about 0.01~mag, and 
astrometry 0.15~arcsec (with possible systematic error up to 0.7~arcsec). 
Color-magnitude diagrams of observed fields are also presented. 

The maps of the SMC are the first from the series of similar maps covering 
other OGLE fields: LMC, Galactic bulge and Galactic disk. The data are very 
well suited for many projects, particularly for the SMC which has been 
neglected photometrically for years. Because of potentially great impact on 
many astrophysical fields we decided to make the SMC data available to the 
astronomical community from the OGLE Internet archive.}{Magellanic Clouds -- 
Surveys -- Catalogs -- Techniques: photometrics}

\Section{Introduction}
The Optical Gravitational Lensing Experiment (OGLE) is a long term observing 
project which started in 1992 as the search for microlensing events in our 
Galaxy with the ultimate goal of providing information about the dark unseen 
matter (Udalski \etal 1992). In 1993 the first ever observed microlensing 
event toward the Galactic bulge was detected. In total about 20 microlensing 
events were found during the first phase of the project which ended in 1995 
(Udalski \etal 1994, Wo{\'z}niak and Szyma{\'n}ski 1998). 

Starting from January 1997 the OGLE project entered its second phase -- 
OGLE-II. With a new, dedicated 1.3-m Warsaw telescope located at the Las 
Campanas Observatory, the observing capabilities of the OGLE project increased 
by a factor of 30 (Udalski, Kubiak and Szyma{\'n}ski 1997). New targets, 
namely the Large and Small Magellanic Clouds, new fields in the Galactic bulge 
and Galactic disk have been added to the list of regularly observed regions of 
the sky. 

After the first year of observations large databases of the observed targets 
have been created and first microlensing events have been detected (Udalski 
and Szyma{\'n}ski 1998). The OGLE project photometric observations are 
collected in the standard {\it BVI}-bands which makes them very well suited 
not only for microlensing but also for many side projects. For instance,  
photometry of the LMC and SMC has already been used for a new distance 
determination to both Magellanic Clouds (Udalski \etal 1998, Udalski 1998). 

In this paper, first of the series, we present the {\it BVI} maps of dense 
stellar regions in the SMC observed in the course of the OGLE-II project. The 
maps provide {\it BVI}-band photometry of all stellar objects detected in 
observed fields and in the case of the SMC contain well calibrated photometry 
and astrometry of about 2.2 million stars from the central regions (${\approx 
2.4}$ square degree) of the SMC bar. The data are used to construct 
color-magnitude diagrams (CMDs) of observed fields revealing many subtle 
features of stellar populations in the SMC. 

Presented maps can be a very useful tool for many astronomical projects. For 
instance, a unique catalog of clusters in the SMC with accurate 
color-magnitude diagrams follows this paper (Pietrzy{\'n}ski \etal 1998). The 
SMC observations with the modern, precise techniques are rare and those which 
can be found in the literature concentrate on particular objects. The maps 
contain a full variety of objects including field stars and clusters located 
in different parts of the SMC and therefore are ideal for comparisons of 
stellar populations in different regions of the SMC. 

Bearing in mind potential impact of the OGLE photometric data on our 
understanding of the SMC and other OGLE objects, we decided to make these data 
available to the astronomical community. At the end of this paper we provide 
information how to obtain the SMC maps. In the next papers of this series the 
LMC and Galactic bulge maps will be gradually released. 

In subsequent Sections we describe observations, reduction and calibration 
techniques used to construct the maps. Then results of accuracy and 
completeness tests are presented. Finally, we show CMDs of all our fields in 
the SMC. 

\Section{Observations}
All observations presented in this paper were carried out with the 1.3-m 
Warsaw telescope at the Las Campanas Observatory, Chile, which is operated by 
the Carnegie Institution of Washington, during the second phase of the OGLE 
experiment. The telescope was equipped with the "first generation" camera with 
the SITe ${2048\times2048}$ CCD detector. The pixel size was 24~$\mu$m 
resulting in 0.417~arcsec/pixel scale. The observations of the SMC were 
performed in the "slow" reading mode of the CCD detector with the gain 
3.8~e$^-$/ADU and readout noise ${\approx5.4}$~e$^-$. Details of the 
instrumentation setup can be found in Udalski, Kubiak and Szyma{\'n}ski 
(1997). 

Observations of the SMC started on June~26, 1997. As the microlensing search 
is planned to last for a few years, observations of selected fields will be 
continued during next few seasons. In this paper we present data collected 
before the end of the first SMC season -- Mar.~4, 1998. Observations were 
carried out in the driftscan mode. In this mode the CCD detector columns are 
aligned along the drift direction and the detector is read continuously with 
the rate synchronized with the drift rate of the star. As a result much larger 
areas can be observed with a single chip than in the normal, still frame mode. 
Also the dead time for detector reading is avoided. The images are rectangular 
with the width equal to the width of the CCD detector and the length 
proportional to the scanning time. For technical reasons -- handling of large 
data files -- OGLE driftscans are limited to 8192 lines which produces images 
of ${2048\times 8192}$ pixel size (${\approx34}$~MB raw image). The OGLE 
driftscans are made with the drift along the meridian (\ie declination) 
direction. In this way any region of the sky is accessible for observations 
including the Magellanic Clouds located at high southern declinations. The 
drift of stars is forced by tracking the telescope in both axes: normal 
tracking in RA and additional tracking in declination. The declination 
tracking rate determines the effective exposure time which in the case of the 
SMC is 125~sec, 174~sec and 237~sec for the {\it I, V} and {\it B}-bands, 
respectively. Very sensitive CCD detector allows to obtain good photometry 
down to ${V\approx21.5}$~mag even with such relatively short exposure times. 

Observations are made in three bands: {\it B,V} and {\it I}. Because of 
microlensing observing strategy, the vast majority of observations is 
collected in the {\it I}-band with some additional measurements in the {\it B} 
and {\it V}-bands. Standard set of glass Schott filters is used, closely 
approximating the standard {\it BVI} system. 

Because of large stellar density in the OGLE-II main targets, the observations 
are limited to very good seeing conditions only, in order to ensure good 
photometry. Typical median seeing of presented data set is about 1.25~arcsec 
in the {\it I}-band with the best images reaching 0.8~arcsec. It should be 
noted that the active driftscans with open loop (no guiding) tracking in two 
axes give always slightly worse seeing because of inaccuracies in tracking 
etc.\ Observations of the SMC are usually stopped when the seeing exceeds 
1.8~arcsec. 

11 driftscan fields were selected in the SMC. They cover practically all dense 
regions of the SMC bar. Fig.~1 presents the picture of the SMC from the 
Digitized Sky Survey CD-ROMs with contours of our fields. Coordinates of the 
center of each field are given in Table~1. Each of the fields covers 
${\approx14.2\times57}$~arcmin on the sky (${\approx0.22}$ square degree). 
Fields are shifted in declination to follow the location of the center of the 
SMC bar on the sky. Neighboring fields overlap by about 1~arcmin for 
calibration of photometry and astrometry purposes. In the next observing 
seasons additional fields will be added to those listed in Table~1 to cover 
practically entire SMC. 
\MakeTable{lcc}{12.5cm}{Equatorial coordinates of the SMC fields}
{
\hline
\noalign{\vskip3pt}
\multicolumn{1}{c}{Field} & RA (J2000)  & DEC (J2000)\\
\hline
\noalign{\vskip3pt}
SMC$\_$SC1  & 0\uph37\upm51\ups & $-73\arcd29\arcm40\arcs$\\
SMC$\_$SC2  & 0\uph40\upm53\ups & $-73\arcd17\arcm30\arcs$\\
SMC$\_$SC3  & 0\uph43\upm58\ups & $-73\arcd12\arcm30\arcs$\\
SMC$\_$SC4  & 0\uph46\upm59\ups & $-73\arcd07\arcm30\arcs$\\
SMC$\_$SC5  & 0\uph50\upm01\ups & $-73\arcd08\arcm45\arcs$\\
SMC$\_$SC6  & 0\uph53\upm01\ups & $-72\arcd58\arcm40\arcs$\\
SMC$\_$SC7  & 0\uph56\upm00\ups & $-72\arcd53\arcm35\arcs$\\
SMC$\_$SC8  & 0\uph58\upm58\ups & $-72\arcd39\arcm30\arcs$\\
SMC$\_$SC9  & 1\uph01\upm55\ups & $-72\arcd32\arcm35\arcs$\\
SMC$\_$SC10 & 1\uph04\upm51\ups & $-72\arcd24\arcm45\arcs$\\
SMC$\_$SC11 & 1\uph07\upm45\ups & $-72\arcd39\arcm30\arcs$\\  
\hline}

During the first observing season about 100--150 observations in the 
{\it I}-band were collected, depending on the field. For two fields, 
SMC$\_$SC5 and SMC$\_$SC6 a few additional measurements were collected 
earlier, in January 1997. Number of {\it B} and {\it V} observations is 
smaller -- about 15 and 30 per field, respectively. It is though, large 
enough to derive precise photometry of objects in these bands. 

\Section{Photometric Reductions}
\Subsection{The OGLE Data Pipeline}
When a new frame is collected it is automatically intercepted by the OGLE 
photometric data reduction pipeline. All reductions are performed almost in 
real time at the telescope. 

In the first step preliminary reductions are performed -- debiasing, and 
flat-fielding based on regularly collected twilight sky flat-field and bias 
images. It should be noted that the driftscan mode flat-fields are 
one-dimensio\-nal line which is usually an average of a few thousand lines 
(depending on the length of the calibration driftscan flat-field image) and 
thus determined very precisely. 

After the frame is de-biased and flat-fielded it is sent to the second stage 
reduction procedures which derive photometry of stars on the image. The 
photometry is calculated with the modified version of the {\sc DoPhot} 
photometry program (Schechter, Saha and Mateo 1993). Because of significant 
variations of the Point Spread Function (PSF) across the image caused by some 
inaccuracies of the telescope tracking and variable seeing in different 
moments of scanning, photometry reductions are performed on the 64 subframes 
of ${512\times 512}$ pixel size on which the PSF can be assumed as constant in 
the first approximation. The {\sc DoPhot} program is run in the so called fixed 
position mode, that is an input list of positions of objects in a given field, 
generated in advance, is provided to the program. Coordinates from this list 
are transformed to fit the coordinates of the current frame and fixed for PSF 
photometry fitting. This mode is much faster and more accurate than the 
regular mode, in particular when the seeing of the frame is worse. 

The input list of coordinates comes from the so called template image which is 
one of the best images of the entire data set for a given field and filter 
taken at the possibly best seeing conditions. Positions of objects from the 
template images are obtained from photometric reductions with a special 
procedure based on regular and fixed mode reductions with the {\sc DoPhot} 
program. Photometry of each subframe of the template image is tied based on 
photometry in overlapping regions between subframes (for template reduction 
somewhat larger subframes are cut from the template image to reach 100
pixel overlap). Photometry of the template image is further treated as 
the reference instrumental photometry for the entire data set. Template 
coordinate lists for the {\it B} and {\it V}-bands are generated in similar 
way with the exception that positions of objects in these frames are fitted 
first with those of the {\it I}-band template image. In this way position of 
the star is always the same in the list of each band photometry which makes 
further data handling much easier. 

When reductions of all subframes of a given image are done, photometry
of each  subframe is tied to that of the reference template subframe and
results of  photometry are stored in compact binary format in disk
files. The process of  reductions: dividing to subframes, reductions,
correction of the zero points  etc.\ is performed fully automatically. The
amount of CPU of the system is  sufficient to perform photometric
reductions of all frames  collected during the night (50--70 driftscans)
within 24 hours.

\Subsection{Transformation to the Standard System} Several Landolt
(1992) fields were observed during about 30 photometric nights  to
derive transformation of our instrumental magnitudes in the template 
photometric system to the standard {\it BVI} system. Based on a few
hundreds  of measurements of standard stars the following mean
transformation was  derived:  
\begin{eqnarray}
B&=&b-0.041\times(B-V)+{\rm const}_B\nonumber\\
V&=&v-0.002\times(V-I)+{\rm const}_V\nonumber\\
I&=&i+0.029\times(V-I)+{\rm const}_I\\ 
B-V&=&0.959\times(b-v)+{\rm const}_{B-V}\nonumber\\ 
V-I&=&0.969\times(v-i)+{\rm const}_{V-I}\nonumber 
\end{eqnarray} 
where lower case letters {\it b,v,i} denote the aperture magnitudes
normalized  to 1~sec exposure time. The transformation coefficients are
the mean for the  entire season and were determined on a few nights when
number of measurements  of standard stars was large. The zero points and
extinction coefficients (when impossible, mean extinction values were
used) were  determined individually for each night. Residuals between
fitted magnitudes of  standard stars and observed ones did not exceed
0.02--0.03~mag. 

As can be seen the transformation coefficients are close to zero for
magnitude  or one for color transformations, thus indicating that our
instrumental system  very well approximates the standard one and
transformation errors are small.  As the first step to convert our
instrumental system to the standard one,  aperture corrections to the
template magnitudes were derived. A special  procedure was written,
based on the {\sc DoPhot} program, which reduces each  subframe of a
given image in analogous way as during normal reductions,  selects
bright stars well separated from neighboring stars which could 
contaminate their photometry and calculates the aperture magnitude for
every  selected star removing all remaining objects from the frame.
About 10--50 aperture stars were usually used in each of the 64
subframes of a given image.  The median of aperture corrections derived
from these stars was adopted as the  aperture correction of a given
subframe. Next, the "total correction" for each  subframe of a given
image was calculated. This is defined as the sum of its aperture
correction, the transformation  zero point, extinction correction and
normalization to 1~sec exposure time. Then "total corrections"
determined from about 25, 10 and 6 nights for  the {\it I, V} and {\it
B}-bands, respectively, were averaged producing  an array of 64 mean
"total corrections" for a given field and band. Typical mean error of
the final "total correction" values was less than 0.01~mag.

\Subsection{Final Databases of Results and BVI Maps of the SMC} "Total
corrections", determined as described in the previous Section, were 
used to convert the template photometric system of each field and filter
to  the instrumental system very close to the standard one (the
difference involves only  color terms of Eqs.~(1)). At this stage
databases of photometric results were  created and the appropriate
"total corrections" were simply added to the  template photometric
system magnitudes when databases were filled up. The  databases used are
very similar to those developed for the OGLE-I phase of the  project
(Szyma{\'n}ski and Udalski 1993) with some slight improvements and 
modifications. They contain all individual photometric measurements in
the  system close to the standard one (no color terms) and all basic
information  about each frame: heliocentric Julian date, exposure time,
mean seeing and its  standard deviation, frame grade etc. Entire database
for a given field and  filter consists of two parts -- that of stars
detected on the template frame  and that of "new" objects. In this paper
we limit ourselves to the template  objects only. 
The databases of the SMC contain approximately 2.25 million objects. The vast 
majority of them are objects classified by the {\sc DoPhot} program as 
stellar. 
\MakeTable{lc}{12.5cm}{Number of stellar objects in the SMC maps}
{\hline
\noalign{\vskip3pt}
\multicolumn{1}{c}{Field} & No of stellar\\
                          &   objects\\
\noalign{\vskip3pt}
\hline
SMC$\_$SC1                   &   120002\\ 
SMC$\_$SC2                   &   107326\\
SMC$\_$SC3                   &   240045\\
SMC$\_$SC4                   &   198201\\
SMC$\_$SC5                   &   319850\\
SMC$\_$SC6                   &   326367\\
SMC$\_$SC7                   &   268006\\
SMC$\_$SC8                   &   211115\\
SMC$\_$SC9                   &   176832\\
SMC$\_$SC10                  &   140589\\
SMC$\_$SC11                  &   120932\\
\hline
}

When the databases of photometric results were created for each field and 
filter for our SMC fields, the final {\it BVI}-band maps of the SMC were 
constructed. First, measurements of each star were averaged with $5\sigma$ 
rejection alghoritm in each band. Then color term transformation corrections 
were calculated and finally the standard system magnitudes were derived. Only 
magnitudes derived from more than 40, 10 and 5 good measurements in the {\it 
I, V} and {\it B}-bands, respectively, were listed in the final maps of the 
SMC. Good measurement was defined as the one with the stellar type object 
returned by the {\sc DoPhot} program and the error returned by {\sc DoPhot} 
not exceeding 1.6 median of the errors from the entire set of measurements of 
a given star. Final number of objects included in the SMC maps is given in 
Table~2.

\begin{figure}[p]
\vspace*{2cm}
\centerline{\hglue8.3cm\psfig{figure=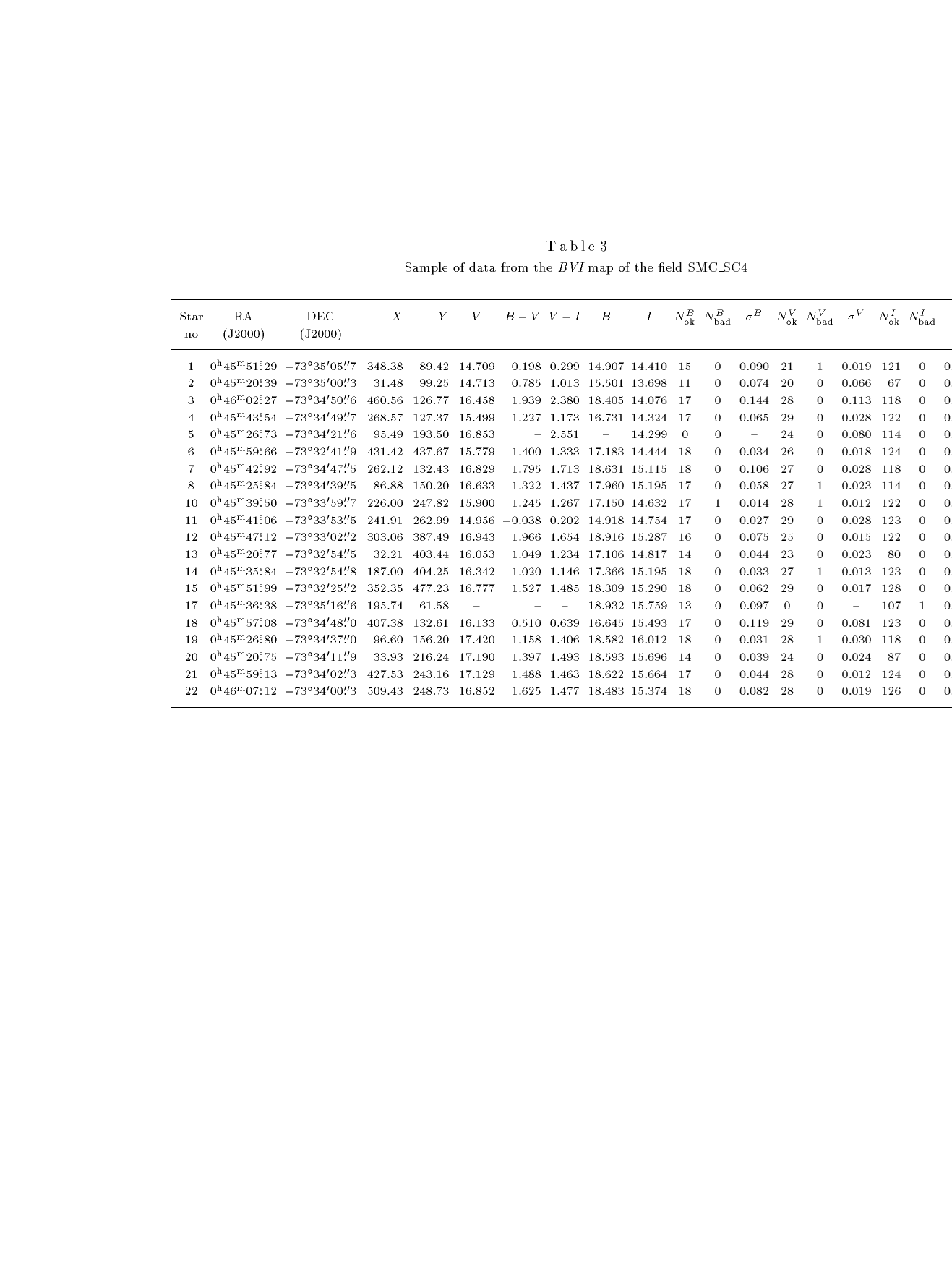,height=23cm,angle=90}}
\end{figure}
Table~3 presents sample data of the final map for the SMC$\_$SC4 field. The 
following data are provided: star number in the database, equatorial 
coordinates (RA and DEC, J2000) of the star, $x$ and $y$ positions in the 
{\it I}-band template image, magnitudes: {\it V}, ${B-V}$, ${V-I}$, {\it B} 
and {\it I}, number of observations, number of rejected observations and 
standard deviation of the mean magnitude for {\it B}, {\it V} and 
{\it I}-bands, respectively. Such maps for all our fields are available to the 
astronomical community (see Section~7). 

\Section{Astrometry}
To convert positions of stars to the equatorial system the pixel coordinate 
system of the {\it I}-band template image for each field was transformed to 
the equatorial system using the Digital Sky Survey (DSS) images. 

The procedure consists of the following steps. First, FITS image, somewhat 
larger than the area of a given field, is extracted from the DSS CD-ROMs. 
Then, the "autofind" program is run on that frame. It finds all objects 
brighter than selected threshold and calculates a centroid of each detected 
object. Now, the main transformation program is run. First, it converts the 
$(x,y)$ coordinates of each star from the DSS image to RA and DEC based on 
prescription and procedures provided with the DSS distribution. Then (RA, DEC) 
of each star are transformed with simple trigonometric formulae to $(x',y')$ 
coordinates of the Cartesian system on the plane tangent to the celestial 
sphere at the center of a given field. Finally, transformation between 
$(x',y')$ coordinates and pixel coordinate system of the {\it I}-band template 
image follows. The transformation between those two systems of coordinates is 
relatively simple and third order polynomials are sufficient to achieve good 
accuracy. The transformation is calculated in a few iterations with decreasing 
critical radius within which objects are treated as identical. The final 
critical radius is 0.75~arcsec. About 3000--8000 stars in the SMC fields were 
used for transformation depending on stellar density of the field. 

\Section{Data Tests}
\Subsection{Photometry}
To asses quality of photometry a few tests were performed. First, we compared 
photometry of the same stars located in the overlapping regions between 
fields. As the overlapping regions extend along the length of the image we 
compared the magnitude differences of stars on both frames as a function of 
line number. Figs.~2, 3 and 4 present plots "magnitude difference \vs CCD line 
number" for fields located in the west, center and east regions of the 
SMC (SMC$\_$1--2, SMC$\_$5--6, and SMC$\_$9--10) for filters {\it B, V} and 
{\it I}. Similar plots for other fields look practically identical. 
As can be seen the photometry in overlapping regions of neighboring fields is 
in very good agreement. The mean difference is always below 0.01~mag 
indicating that our calibration procedures were correct. 

In the second test we compared our photometry with other reliable measurements 
from the literature. Unfortunately good quality observations of the SMC with 
modern techniques are rare. We limited our comparison mostly to CCD 
measurements, as only those can be reliable in crowded fields of the SMC. We 
found the following past observations of the SMC: 

\begin{itemize}
\itemsep=-3pt
\item CCD photometry of NGC~346 was published by Massey, Parker and Garmany 
(1989). Only part of this cluster is located in the SMC$\_$SC8 field. The mean 
difference between the OGLE and Massey, Parker and Garmany (1989) photometry is 
${\Delta V{=}0.003{\pm}0.032}$~mag (42 stars) and $\Delta(B{-}V){=}{-
}0.022{\pm}0.033$ mag (31 stars). 

\item CCD photometry of NGC~330 (SMC$\_$SC7 field) was
obtained  by Walker (Caloi \etal 1993). Photometry of Walker is in 
good agreement with more recent photometry of Vallenari, Ortolani and
Chiosi (1994). The mean difference between the OGLE and Walker photometry
is $\Delta V=-0.002\pm0.020$ mag and
${\Delta(B-V)=0.040{\pm}0.014}$~mag (9 stars). 

\item CCD photometry of the field around HV~1876 (SMC$\_$SC9 field) variable 
star (Jensen, Clausen and Gim\'enez 1988). The mean difference between the OGLE 
and Jensen, Clausen and Gim\'enez (1988) photometry of proposed standard stars 
is ${\Delta V=0.004\pm0.030}$~mag and $\Delta(B-V)=0.002\pm0.018$ mag. 

\item CCD photometry of the field around HV~2016 (SMC$\_$SC11 field) variable 
star (Jensen, Clausen and Gim\'enez 1988). The mean difference between the OGLE 
and Jensen, Clausen and Gim\'enez (1988) photometry of proposed standard stars 
is ${\Delta V=-0.007\pm0.011}$~mag and $\Delta(B-V)=0.015\pm0.006$ mag. 

\item Photoelectric photometry sequence around NGC~419 (SMC$\_$SC11
field) was  obtained by Wenderoth \etal (1994). Although we do not trust
photoelectric  photometry in such dense fields because of crowding we
checked photometry of the  brightest stars of that sequence which are
still not saturated in our images. The mean  difference between the OGLE and
Wenderoth \etal (1994) photometry is $\Delta V=- 0.007\pm0.010$~mag,
$\Delta(B-V)=-0.059\pm0.013$ mag and $\Delta(V-I)=-
0.001\pm0.032$~mag 

\item Finally we compared the {\it I}-band light curves of two "stable" 
variable stars from the NGC~330 region (SMC$\_$SC7 field), namely Cepheid 
variables \#389 and \#952, observed by Sebo and Wood (1994) to those 
retrieved from the OGLE databases -- Fig.~5. It can be seen that zero points 
of both photometries are in very good agreement confirming that OGLE zero 
points are determined correctly. 
\end{itemize}

Summarizing, very good agreement of photometry derived independently in 
neighboring fields and comparison with reliable measurements from the 
literature indicate that our photometry is calibrated correctly. We estimate 
that the systematic error of our photometry should not exceed 0.01~mag. 

\Subsection{Completeness}

To asses completeness of our three color maps of  the SMC we performed a
series of  artificial stars tests.  We selected for tests $512\times
512$ pixel subframes of each band template images of three fields. These
fields are located in the east, center and west regions of the observed
part of the SMC, namely SMC$\_$SC1, SMC$\_$SC6 and SMC$\_$SC10 and
represent the least and most dense regions of the SMC bar. Selected
subframes cover the central, most dense part of each field. 

To each subframe a number of  artificial stars in randomly selected
positions was added using a special software based on {\sc DoPhot}
procedures. In order not to change  crowding of the frame we added only
296 stars in each test with magnitude distribution as shown in column 1 of 
Table~4. Then we checked how many of them were recovered with our standard 
reduction pipeline treating the input and output star as identical if its
difference of position and magnitude was smaller than 0.8~arcsec and 
$\pm0.5$~mag, respectively. One hundred such tests were  performed for each 
subframe. The rate of recovered objects is listed in Table~4. 

\setcounter{table}{3}
\MakeTable{c@{\hspace{6pt}}c@{\hspace{6pt}}c@{\hspace{6pt}}c@{\hspace{6pt}}c
@{\hspace{6pt}}c@{\hspace{6pt}}c@{\hspace{6pt}}c@{\hspace{6pt}}c
@{\hspace{6pt}}c@{\hspace{6pt}}c@{\hspace{6pt}}c@{\hspace{6pt}}c}
{12.5cm}{Completeness of the SMC maps}
{
\hline
Stars &&\multicolumn{3}{c}{Completeness}&&\multicolumn{3}{c}{Completeness}&&\multicolumn{3}{c}{Completeness}\\
per bin & $B$  & SC$\_$1 &  SC$\_$6 &  SC$\_$10 & $V$  & SC$\_$1 &  SC$\_$6 &  SC$\_$10 & $I$  & SC$\_$1 &  SC$\_$6 &  SC$\_$10 \\
\hline
~2 & 14.0 & 100.0 & 100.0 & 100.0 & 14.5 & 100.0 &  99.0 & 100.0 & 13.8 & 100.0 & 100.0 & 100.0 \\ 
~5 & 14.5 & 100.0 &  99.6 & 100.0 & 15.0 &  99.6 & 100.0 &  99.8 & 14.3 &  99.4 & 100.0 &  99.8 \\ 
~7 & 15.0 &  99.4 &  99.7 &  99.9 & 15.5 &  99.9 &  99.4 & 100.0 & 14.8 & 100.0 &  99.9 &  99.9 \\ 
10 & 15.5 &  99.6 &  99.3 &  99.7 & 16.0 & 100.0 &  99.3 &  99.3 & 15.3 &  99.8 &  99.4 &  99.9 \\ 
12 & 16.0 &  99.0 &  99.7 &  99.7 & 16.5 &  99.6 &  99.3 &  99.5 & 15.8 &  99.6 &  99.3 &  99.5 \\ 
15 & 16.5 &  98.7 &  98.8 &  99.8 & 17.0 &  99.1 &  98.5 &  99.3 & 16.3 &  99.9 &  99.5 &  99.8 \\ 
17 & 17.0 &  98.8 &  98.4 &  99.0 & 17.5 &  98.8 &  98.6 &  99.1 & 16.8 &  99.5 &  99.1 &  98.9 \\ 
20 & 17.5 &  97.9 &  98.4 &  99.7 & 18.0 &  98.8 &  97.8 &  98.7 & 17.3 &  99.5 &  98.8 &  99.1 \\ 
22 & 18.0 &  98.1 &  96.7 &  98.7 & 18.5 &  98.0 &  96.0 &  97.3 & 17.8 &  98.7 &  97.5 &  98.4 \\ 
25 & 18.5 &  97.2 &  94.5 &  98.2 & 19.0 &  97.2 &  93.0 &  96.1 & 18.3 &  98.5 &  95.1 &  98.0 \\ 
27 & 19.0 &  97.1 &  90.9 &  96.7 & 19.5 &  94.4 &  90.3 &  95.0 & 18.8 &  98.2 &  93.6 &  96.4 \\ 
30 & 19.5 &  94.7 &  83.7 &  95.0 & 20.0 &  93.5 &  86.3 &  92.0 & 19.3 &  96.8 &  89.3 &  95.9 \\ 
32 & 20.0 &  91.7 &  74.8 &  89.1 & 20.5 &  89.5 &  78.9 &  87.2 & 19.8 &  95.2 &  84.8 &  93.2 \\ 
35 & 20.5 &  87.0 &  58.9 &  79.6 & 21.0 &  78.8 &  65.0 &  76.0 & 20.3 &  91.8 &  72.4 &  85.8 \\ 
37 & 21.0 &  77.9 &  37.4 &  33.1 & 21.5 &  53.0 &  42.9 &  56.4 & 20.8 &  76.7 &  42.1 &  55.3 \\ 
\hline
}

It is clear  that down to ${B\approx20.0}$, ${V\approx20.5}$ and
${I\approx20.0}$ detection completeness is very high and then it fades 
gradually reaching $\approx 50\%$ at ${B\approx21.2}$, ${V\approx21.5}$
and ${I\approx21.0}$ for the least dense fields. This figures might be
assumed as limiting magnitudes of our survey. As expected completeness
is somewhat lower for the most dense field SMC$\_$SC6.

Our maps contain only objects which were detected in the {\it  I}-band
templates. Therefore the {\it I}-band completeness tests correspond to
completeness of our maps.  Additional objects present only in {\it V}
and/or {\it B}-band templates as  well as objects from "new" databases
were not included in the maps. We  estimate that only small percentage
of objects was omitted. 

\Subsection{Astrometry}
Accuracy of our astrometric solutions was checked by comparison of
coordinates  of detected objects located in overlapping regions between
fields. First,  transformation between fields was derived and common
objects were identified.  Then, equatorial coordinates of each
identified object were calculated based  on transformations for both
fields and the difference between the two. Finally, the  mean difference
and its standard error were calculated from the entire sample  of
identified pairs of stars (typically a few hundred objects). The mean 
difference of coordinates between fields was typically about
0.1--0.15~arcsec  with the standard deviation of about 0.1--0.2~arcsec.
This error is  representative of internal error of our determination of
equatorial  coordinates. 

To check possible systematic errors we compared fields with astrometric 
solutions based on different plates from the DSS. We repeated the above
procedure  comparing coordinates of fields determined from different 
plates of the DSS. In this case the mean difference was much larger,
sometimes reaching 0.7~arcsec. This number might be assumed as the
possible systematic  error of the DSS coordinates system and thus of our
equatorial coordinates. 

\Section{Color-Magnitude Diagrams}
Based on our three color maps of the SMC, we constructed composite 
color-magnitude diagrams for each of our fields. Figs.~6--16 present the 
${V{-}(B{-}V)}$ and ${V-(V-I)}$ CMDs of all our fields in the SMC. Only 
10--30\% of stars were included in these Figures for clarity.

The CMDs of the SMC fields reveal in great detail main features known from  
previous studies: the main sequence, red giant branch with nicely pronounced  
red clump of giants, its vertical extension (similar to that observed in the 
LMC, \cf Zaritsky  and Lin 1997, Beaulieu and Sackett 1998, Udalski \etal 
1998) and asymptotic  branch clump. Also many other fine features become 
recognizable due to large  statistics of stars in these diagrams. It is 
obvious that our maps provide a  unique tool for analysis of the wide variety 
of topics concerning the SMC and  stars in general. 

\Section{Conclusions}
We release in this paper the three color, {\it BVI}-band maps of the SMC. The 
data contain photometry for about 2.2 million stars from the central parts of 
the SMC bar. The data are precisely calibrated, the mean magnitudes are 
determined from tens/hundreds of individual measurements. Our tests show that 
accuracy of the zero point of our photometry should be about 0.01~mag. Precise 
equatorial coordinates for all objects are determined from the DSS astrometric 
solution with internal accuracy of about 0.15~arcsec, and possible systematic 
errors of the DSS astrometric system up to 0.7~arcsec. Large number of 
photometric measurements used for determination of mean magnitudes makes our 
unique data a huge set of secondary photometric standards in the SMC. 

In the next paper of this series similar data for the LMC bar fields observed 
during the OGLE-II project will be presented. Release of 
similar maps of the Galactic bulge is planned in the future. 

\newpage
The three color maps are available to the astronomical community from
the OGLE  archive {\it http://www.astrouw.edu.pl/\~{}ftp/ogle} and its
mirror\newline  {\it http://www.astro.princeton.edu/\~{}ogle} (or {\it 
ftp://astro.princeton.edu/ogle/ogle2/maps/smc}). Also {\it
I}-band FITS template images of our  fields are included. Due to large
volume of data -- about 400 MB, which is not  always suitable for network
transfers we also distribute the entire set of  data on CD-ROMs for a
small fee covering handling costs. Please contact {\it  via} e-mail: {\it
cfpa@sirius.astrouw.edu.pl}. Usage of the data is allowed under  the
condition of proper acknowledgment to the OGLE project. 

We provide these data in the most original form to avoid any additional 
biases. For instance we do not mask bright stars which often produce
many  artifacts, but such masking could potentially remove some
interesting  information on objects located close to bright stars. We do
not remove objects  which are located in overlapping areas between the
neighboring fields.  Cross-identification of these objects can be easily
done based on provided  equatorial coordinates. 

\vspace*{-6pt}
\Acknow{We would like to thank Prof.\ Bohdan Paczy\'nski for many encouraging 
discussions and help at all stages of the OGLE project. The paper was partly 
supported by the Polish KBN grant 2P03D00814 to A.\ Udalski. Partial support 
for the OGLE project was provided with the NSF grant AST-9530478 to 
B.~Paczy\'nski. We acknowledge usage of The Digitized Sky Survey which was
produced at the Space Telescope Science Institute based on photographic
data obtained using The UK Schmidt Telescope, operated by the Royal
Observatory Edinburgh.}

\newpage
\centerline{\Large\bf Figure Captions}
\vskip1cm
\noindent
Fig.~1. OGLE-II fields in the SMC. North is up and East to the left.

\noindent
Fig.~2. Difference of magnitude in overlapping regions of fields SMC$\_$SC2 
and SMC$\_$SC1.

\noindent
Fig.~3. Difference of magnitude in overlapping regions of fields SMC$\_$SC6 
and SMC$\_$SC5.

\noindent
Fig.~4. Difference of magnitude in overlapping regions of fields SMC$\_$SC10 
and SMC$\_$SC9.

\noindent
Fig.~5. Comparison of photometry of two Cepheid variables observed by Sebo 
and Wood (1994) and OGLE-II.

\noindent
Fig.~6. Color-magnitude diagrams of the SMC$\_$SC1 field. About 20\% stars 
from this field are plotted.

\noindent
Fig.~7. Color-magnitude diagrams of the SMC$\_$SC2 field. About 30\% stars 
from this field are plotted.

\noindent
Fig.~8. Color-magnitude diagrams of the SMC$\_$SC3 field. About 13\% stars 
from this field are plotted.

\noindent
Fig.~9. Color-magnitude diagrams of the SMC$\_$SC4 field. About 13\% stars 
from this field are plotted.

\noindent
Fig.~10. Color-magnitude diagrams of the SMC$\_$SC5 field. About 10\% stars 
from this field are plotted.

\noindent
Fig.~11. Color-magnitude diagrams of the SMC$\_$SC6 field. About 10\% stars 
from this field are plotted.

\noindent
Fig.~12. Color-magnitude diagrams of the SMC$\_$SC7 field. About 13\% stars 
from this field are plotted.

\noindent
Fig.~13. Color-magnitude diagrams of the SMC$\_$SC8 field. About 20\% stars 
from this field are plotted.

\noindent
Fig.~14. Color-magnitude diagrams of the SMC$\_$SC9 field. About 20\% stars 
from this field are plotted.

\noindent
Fig.~15. Color-magnitude diagrams of the SMC$\_$SC10 field. About 20\% stars 
from this field are plotted.

\noindent
Fig.~16. Color-magnitude diagrams of the SMC$\_$SC11 field. About 30\% stars 
from this field are plotted.
\end{document}